\documentclass[12pt]{article}
\usepackage{aaspp4}
\begin{document}

\title{Constraints on Photometric Calibration from Observations
of High-Redshift Type Ia Supernovae}
\author{David W. Hogg\altaffilmark{1,2}}
\affil{Institute for Advanced Study, Olden Lane, Princeton NJ 08540}
\altaffiltext{1}{Hubble Fellow}
\altaffiltext{2}{{\tt hogg@ias.edu}}

\begin{abstract}
The good match of the type Ia supernova (SNIa) Hubble Diagram to the
prediction of a not-unreasonable cosmological world model shows that
measurements of standard stars and their comparison with point sources
down to $m=25~{\rm mag}$ is good to better than $\pm 0.5~{\rm mag}$
over an $11~{\rm mag}$ range.  It also shows that the true spectral
energy distribution (SED) shapes of standard stars are known to better
than $\pm 0.5~{\rm mag}$ over an octave in wavelength.  On the other
hand, the SNIa argument for an accelerating Universe assumes that the
magnitude system is good to much better than $\sim 0.1~{\rm mag}$ over
the $11~{\rm mag}$ range, and that SED shapes are known to much better
than $\sim 10~{\rm percent}$ over an octave in wavelength.  There is
no independent empirical evidence for these plausible assumptions.
\end{abstract}


\section{The type Ia supernovae results}

The type Ia supernova (SNIa) Hubble Diagrams established by the
high-redshift SNIa search and photometry teams (Perlmutter et al 1997;
Garnavich et al 1998; Schmidt et al 1998; Riess et al 1998; Perlmutter
et al 1999) stand as a great astronomical achievement of this decade.
These studies provide a tremendous confirmation of the expanding
Universe and big-bang cosmology.  Along with massive searches for
microlensing events (eg, Alcock et al 1998; Beaulieu et al 1995;
Udalski et al 1994), they show that large, coordinated surveys can be
established to routinely make discoveries and follow them up
uniformly.  Along with soon-to-be completed studies of the cosmic
background radiation, they hold the promise of making direct, precise
measurements of the Universe's kinematics.  In fact, at the time of
writing, the SNIa results already favor an accelerating Universe, eg,
with $(\Omega_M,\Omega_{\Lambda})\approx (0.3,0.7)$ (Riess et al 1998;
Perlmutter et al 1999).

One widely overlooked conclusion which can be drawn from the SNIa
results is that astronomical photometric calibration systems and
techniques are basically correct.  In order to make a precise
cosmological measurement, the SNIa must span many magnitudes in flux.
This requires that it be possible to measure, at the few-percent
level, the relative flux between two sources separated by four orders
of magnitude.  Experiments with this kind of dynamic range are
notoriously difficult in any field of study, but particularly in
astronomy, where very different instrumentation, techniques, and
sources of experimental error become relevant at different magnitude
levels.  Furthermore, because the SNIa must also span a large redshift
interval, different SNIa are observed in different rest-frame
bandpasses.  This requires that the absolute spectral energy
distribution (SED) shapes of the standard stars be known to better
than the accuracy of the SNIa measurements (by a factor of at least
$\sqrt{N}$).

Given the tremendous care with which our photometric standards have
been established and studied, it may not be surprising that the SNIa
results are so good.  However, it is important to note that the SNIa
provide a crucial independent and qualitatively different approach to
calibrating photometric measurements.  Before the SNIa were
established as standard (or standardizable; eg, Riess et al 1996)
candles, and before surveys for them spanned the magnitude and
redshift ranges they currently span, there were no precise tests of
the photometric system by any technique fundamentally different from
those by which the system was initially constructed.  No astronomical
results are secure until they are independently confirmed by
qualitatively different techniques.

The subject of this manuscript is the quantitative constraints placed
by the SNIa results on photometric calibration.

\section{Type Ia supernovae as standard stars}

The standard star system currently spans roughly $0<V<16~{\rm mag}$
(Johnson \& Morgan 1953; Kron et al 1953; Landolt 1973, 1983, 1992).
The system is constructed by performing relative observations of
groups of stars spanning small overlapping magnitude ranges (typically
$\sim 5~{\rm mag}$ each) at successively fainter magnitudes.  The
ranges are reconciled with one another to create the $\sim 16~{\rm
mag}$ range currently in use.  This has created a ``magnitude
ladder,'' with some analogy to the distance ladder, where very faint
standards are tied to slightly brighter standards, which are tied in
turn to brighter still.  It is possible for systematic error to creep
in.  Of course most of the SNIa are much fainter than the end of the
magnitude ladder; there are more opportunities for systematic errors
in measuring relative fluxes between the $\sim 15~{\rm mag}$ standard
stars and SNIa as faint as $25~{\rm mag}$.

In principle the standard star ladder could be made irrelevant to the
SNIa projects if all SNIa were compared with the same few faint
standard stars.  In practice, unfortunately, the brightest SNIa were
compared with brighter standards, because the fainter standards had
not been established.  This dependence on brighter standards will
become less important when new, bright SNIa are discovered, as long as
the new SNIa are compared with the faint standards used with the faint
SNIa.

Possible sources for systematic photometry errors, in the standard
star system or in the comparison of SNIa with standards, include:

{\bf Detector linearity} Detector linearity is generally well
established for both the photomultiplier tubes employed in calibration
and the CCDs employed in SNIa studies, so this is not expected to have
a big effect.  On the other hand, many CCDs (including the
well-studied CCDs in the HST/WFPC2 instrument; Stetson 1998, Whitmore
et al 1999) show a ``charge transfer efficiency'' problem which leads
to flux underestimation which itself is a monotonic function of flux.
This is exactly the kind of bias which could tilt the flux ladder at
the very faint end, although it is only a problem at the few-percent
level in HST/WFPC2 and will typically be an even smaller effect in
high-background ground-based observations, even with similar
instrumentation.

{\bf Exposure time differences} Generally the SNIa are measured with
different exposure times than the standard stars in the SNIa studies
(in part to avoid saturation); also bright standard stars are measured
with different exposure times than faint ones; it is possible that
there are biases in camera shutter controls.  This is probably well
calibrated for most instruments, at the few-percent level or better.
(Also, exposure time changes affect the relative contributions of dark
current, read noise, and sky counts in the image; it is not clear that
such changes naturally lead to systematic errors.)

{\bf Beam switching} Standard star calibration measurements which
include differencing of on- and off-source counts require that the
off-source fields for faint standards be ``cleaner'' than those for
bright standards.  In imaging data, such problems are not likely to be
bigger than the inverse signal-to-noise ratio ($S/N$) at which the
standards are taken, since that is the level at which nearby faint
companions can be observed.  This problem therefore ought to be no
worse than a few percent per $\sim 5$~mag range.

{\bf Angular correlations of stars} Stars are correlated on the sky,
and this correlation will no doubt depend on stellar type and
magnitude.  These correlations could lead to biases in photometric
measurements from the very faint stars correlated with their brighter
neighboring standard star.  Again, this is a problem proportional to
the inverse $S/N$ and therefore ought not to be worse than a few
percent per $\sim 5$~mag range.  This is not a problem at all if SNIa
projects use exactly the same focal-plane aperture as those used in
the standard-star calibration programs.

{\bf Image combination} There can be up to tens-of-percent biases
introduced in photometry when multiple images are combined by median
filtering or averaging with sigma-clipping (eg, Steidel \& Hamilton
1993).

{\bf Difference imaging} SNIa tend to be observed in time-separated
difference images (ie, with and without the SNIa) whereas the
standards tend to be observed in on- and off-source difference images.
Some sources of noise are very different in these two different kinds
of difference, including time variability in the detector and sky for
the former, and the numbers and locations of background sources in the
latter.  Many CCD cameras have few-percent sensitivity variations with
temperature and time.

{\bf Sky brightness} Standard stars tend to be taken at the beginning
and end of the night, SNIa during the darkest hours.  This changes the
relative contributions of dark current, read noise and sky counts to
the images.  Of course it is not clear that such changes naturally
lead to biases.  (However, extinction changes which evolve over the
night can lead to scatter, if not biases, when the standards are not
interleaved into the observing program.)

{\bf Bandpasses} Filters of the same name on different detectors at
different telescopes will be at least slightly different.  This can
lead to color terms in the photometric systems established with one
detector but used to study SNIa with another.  The simple fact that
the slopes of the sensitivity-wavelength relationships are different
at the tens of percent level for different detectors will lead to
few-percent differences in broad bandpasses even when identical
filters are employed.

{\bf Clouds and atmosphere} SNIa measurements may be made with less,
or at any rate different, attention paid to atmospheric conditions
than the standard star calibration measurements.  Furthermore, SNIa
measurements and standard star calibration have been done at different
sites.  Even at a fixed site, extinction coefficients for different
bandpasses vary with time by factors of a few, and change in color
(Landolt 1992).  These color changes will affect the shape of the
total throughput, telescope plus atmosphere, at the ten-percent level;
it will affect relative calibration only at the few-percent level,
because standard stars and SNIa are compared through the same
bandpass.  The magnitude of the problem depends on the differences
between the SED shapes of the SNIa and the standards.

{\bf Signal-to-noise} SNIa, comparison standards, and the stars in the
magnitude ladder are all measured at different $S/N$; some biases
depend on $S/N$ alone (Hogg \& Turner 1998).  These are proportional
to inverse $S/N$; they can only affect the very faintest SNIa at the
five to ten-percent level.

It is not clear that any of these possible sources of systematic error
will in fact be significant.  However, there are enough of them that
it is a testament to the care of those who build and calibrate
instruments, calibrate the photometric system, and collect and study
SNIa that the listed effects do not ruin the SNIa Hubble Diagram.

In fact, the SNIa Hubble Diagram is consistent with a set of
cosmological world models within the reasonable range $0<\Omega_M<1$
and $0<\Omega_{\Lambda}<1$ (Riess et al 1998; Perlmutter et al
1999).  Since this reasonable range spans a magnitude difference of
$\pm 0.5~{\rm mag}$ (when tied down to the fluxes of the low-redshift
SNIa), the SNIa Hubble Diagram constrains the drift or systematic
error in the magnitude system to be less than $\pm 0.5~{\rm mag}$ over
$11~{\rm mag}$, or less than $0.045~{\rm mag}$ per magnitude.

If the accumulated systematic error is treated as a tilt in the
magnitude vs log flux diagram, the SNIa constraint corresponds to the
statement that magnitude $m$ is related to flux $f$ by $m=(-2.50\pm
0.11)\log_{10}f+C$.  Constraining systematic error functions more
complicated than a linear tilt is difficult with the current sample of
known SNIa, which has very few in the redshift range $0.1<z<0.4$.
This range is crucial for investigating the magnitude-dependence of
any systematic errors, since it spans a large range in magnitude but
is not strongly affected by changes in the world model.

It is possible to remove the world model uncertainty by just
considering the $\sim 6$~mag range of $z<0.1$ SNIa observations whose
interpretation is relatively independent of cosmological world model.
Although the interpretation of these SNIa has less dependence on world
model, the constraint on the photometric system is weaker, because the
magnitude baseline is shorter.

A similar constraint on the magnitude system can be derived from
photometry of Cepheids in the water-maser galaxy NGC~4258 (Maoz et al
1999), where the absolute distance is known from the kinematics of the
water masers near the nucleus of the galaxy (Hernstein et al 1999).
The comparison with the Large Magellanic Cloud spans 11~mag, and the
uncertainty, including both the NGC~4258 and LMC distance
uncertainties, is on the order of $0.3$~mag.

Taken at face value, the SNIa results currently favor an accelerating
Universe with $\Omega_{\Lambda}>0$.  In the SNIa Hubble Diagram, these
world models are separated from non-accelerating world models with
$\Omega_{\Lambda}=0$ by only $\approx 0.1~{\rm mag}$.  (At fixed
$\Omega_{M}$, accelerating and non-accelerating world models are
separated by more than $0.1~{\rm mag}$.  However, the closest
non-accelerating world model to any non-accelerating one is as close
as $\approx 0.1~{\rm mag}$.)  Until there is independent empirical
evidence that relative photometry techniques are linear to much better
than $0.1~{\rm mag}$ over that $11$~mag range, the SNIa will not
particularly favor accelerating ($\Omega_{\Lambda}>0$) world models
over non-accelerating ($\Omega_{\Lambda}=0$) ones.

\section{Type Ia supernovae as SED-shape calibrators}

Observations of SNIa currently span much of the redshift range
$0<z<1$, so observations in a particular wavelength bandpass span a
range of emitted wavelengths.  For this reason, even if the underlying
SED shapes of SNIa are unknown, the mere fact that they are standard
(or standardizable) candles implies that they can be used to calibrate
the relative sensitivities of different bandpasses.

Usually observations are carried out all in a particular, fixed set of
observational bandpasses, so the magnitudes must be k-corrected.  The
k-correction is the difference between the observed magnitude of a
redshifted source and the magnitude which would have been observed for
the source at the same distance but zero redshift.  It depends on the
individual SED shape of the source being observed because it is a
logarithmic ratio of absolute fluxes in different bandpasses (observed
and emitted).  Here, clearly the k-correction is as good as our
knowledge of the SED shape of the source.  Although the source can be
compared very accurately to standard stars such as Vega, the SED shape
can only be known as well as the SED shapes of the standard stars.

In principle a SNIa project could be designed such that sources at
different redshifts are observed in different bandpasses, matched so
that the observed fluxes of the SNe are observed at the same emitted
wavelengths at all redshifts.  This technique is also dependent on the
SED shapes of the standard stars, because SNIa at different redshifts
will have to be calibrated against different parts of the standard
stars' SEDs.

The SED shapes of Vega and other standard stars are measured by
comparison with laboratory blackbodies of known temperatures.  The
blackbody is close to the telescope (relative to the standard stars!),
so airmass corrections have to be extrapolated from zero airmass to
the airmasses of the stellar observations (Hayes 1970; Oke \& Schild
1970).  An alternative method of absolute calibration makes use of
synthetic photometry of model stellar atmospheres (eg, Colina \&
Bohlin, 1994).

Possible sources for systematic errors in standard star SEDs include:

{\bf Laboratory blackbody temperatures} The inferred SED shapes are
really relative to the laboratory blackbody SED shape, so errors in
temperature lead to SED shape errors.  However, the laboratory
blackbodies are very precise, so there is unlikely to be much
temperature uncertainty; certainly $\Delta T/T<10^{-3}$ (Hayes 1970;
Oke \& Schild 1970).

{\bf Illumination geometry} The blackbodies are point sources near the
telescope, calibrated in luminosity, whereas stars are point sources
at infinity and are being calibrated in flux.  The two will not
illuminate the telescope and its instrumentation identically.  The
experiments are done carefully, so this error is not likely to be
bigger than the angle the telescope aperture subtends to the
blackbody, or on the order of a few percent.

{\bf Absorption layers in the atmosphere} The extrapolation of the
blackbody observations from zero to finite airmass depends on an
extrapolation of airmass corrections from observations at airmasses
of, say, 1 to 2 down to zero.  This extrapolation is not trivial if
there are non-uniform absorbing layers in the atmosphere, or if the
absorption at some wavelengths happens mainly at low altitude.  The
extrapolation has been tested at the ten-percent level (Stebbins \&
Kron 1964).

{\bf Deviations of bandpass shapes} The SNIa and standard stars are
compared in finite bandpasses, not through spectrophotometry.  If any
aspect of bandpass estimation (telescope optics transmission, detector
efficiency, filter curve) is uncertain, an uncertainty is introduced
into the locations and widths of the bandpasses in wavelength space.
This problem is not likely to be big for the SNIa projects, which have
gone to great pains to assess their photometric systems (eg, Kim et al
1996).

{\bf Atmospheric extinction variations} Extinction coefficients for
different bandpasses vary with time by factors of a few, and change in
color, even at a fixed telescope site (Landolt 1992).  These color
changes will affect the shape of the total throughput, telescope plus
atmosphere, at the ten-percent level; it will affect SED-shape
inference at the few-percent level.

{\bf Incorrect model spectra} In the case of synthetic photometric
calibration, the accuracy of the result is directly related to the
accuracy of the model spectra.  This is hard to assess, since the only
calibration-independent tests of model spectra are the equivalent
widths of lines and fractional strengths of spectral breaks, while it
is the absolute level of the continuum that is involved in the
calibration.  However, there are some astronomical sources which are
thought to be very accurately modeled astrophysically.  Synthetic and
blackbody calibrations may disagree at the five-percent level
(eg, Colina \& Bohlin, 1994).

Again, it is not clear that any of these possible sources of
systematic error is significant, but it is nonetheless impressive that
these effects do not ruin the SNIa Hubble Diagram.

The fact that the SNIa Hubble Diagram is consistent with a set of
cosmological world models within the reasonable range $0<\Omega_M<1$
and $0<\Omega_{\Lambda}<1$ constrains the SED error to be less than
$\pm 0.5~{\rm mag}$ over the wavelength range spanned by the redshift
range $0<z<1$, ie, over a factor of two in wavelength.  If SED shapes
could be off by a significant fraction of 10~percent over the factor
of two in wavelength, then the SNIa do not particularly favor
accelerating world models over non-accelerating ones.

\section{Conclusions}

The reasonableness of the SNIa results show that relative photometric
calibration is good to within $\pm 0.5~{\rm mag}$ over $\sim 11~{\rm
mag}$ and that the SED shapes of standard stars are known to $\pm
0.5~{\rm mag}$ over a factor of two in wavelength.  Although perhaps
these constraints are not surprising, they testify to the quality of
the photometric calibration, both of the standard star system, and of
the SNIa projects.  These constraints are important because they are
completely independent of the astronomical techniques used to
construct the calibration in the first place.  If the calibration is
uncertain at the few to ten-percent level over the same magnitude or
wavelength range, then there is no more SNIa evidence for an
accelerating Universe.

Standard candles provide an invaluable resource for testing or,
perhaps, in the future, even establishing systems of calibration.
Unfortunately, they are rare.  However, it is conceivable that certain
kinds of calibration verification similar to that described here could
be performed with massive, uniform sky surveys such as the Sloan
Digital Sky Survey (SDSS).  Because the SDSS collects uniform data on
a huge range of galaxies over a range of redshifts, it will be
possible to constrain certain aspects of photometric calibration.  For
example, if the $r$-band absolute calibration was low by ten percent,
then all populations of extragalactic objects would appear to brighten
at rest-frame 7000~\AA\ in going from redshift $z=0.4$ to $z=0.0$ but
fade at rest-frame 5000~\AA\ over the same redshift interval.
Intercomparison of the evolutionary behaviors of different
extragalactic populations may therefore constrain many aspects of
calibration.  Like SNIa constraints on calibration, these would also
be independent of the standard star system.  This future project
stands as possibly the least glamorous goal of the SDSS.

Unfortunately, the prospects for finding new alternatives for
independent verification of photometric calibration are not good.  The
main approach to improving relative photometry is, and should be,
increased testing of detector and instrument linearity and
repeatability, and continued calibration of standards at fainter
levels and higher signal-to-noise ratio.  Some tests of telescope
linearity could involve ``stopping down'' a large telescope, perhaps
with a randomly perforated entrance cover (since any neutral-density
filter is as hard to calibrate as the photometry itself!).  A
stopped-down telescope would permit some differential tests of
photometry that remove many (though not all) of the aforementioned
systematic problems.  A radical idea, fraught with a new set of
observational difficulties, is to observe, at aphelion and perihelion,
asteroids on highly elliptical orbits around the Sun
(B. Paczynski. private communication).

As for constraining cosmological world models, the SNIa projects will
become much less sensitive to photometric calibration as they push to
higher redshifts, where differing world models make very different
predictions, which are themselves different from the expected
flux-dependence of most of the possible systematic errors.

\acknowledgements It is a pleasure to thank the astrophysicists at the
Institute for Advanced Study in 1999, especially Daniel Eisenstein and
John Bahcall, for lunchtime discussions which culminated in this
study.  Comments from Alex Filippenko, John Gizis, Jim Gunn, Gerry
Neugebauer, Jeff Newman, Bev Oke, Bohdan Paczynski, Jim Peebles,
Michael Richmond, Adam Riess, Tom Soifer and Steve Thorsett were also
extremely helpful.  Support was provided by Hubble Fellowship grant
HF-01093.01-97A from STScI, which is operated by AURA under NASA
contract NAS~5-26555.  This research made use of the NASA ADS Abstract
Service.

\end{document}